# 5.2 Next-Generation Earth System Models: Towards Reliable Hybrid Models for Weather and Climate Applications

**Tom Beucler[1,2], Erwan Koch[2], Sven Kotlarski[3], David Leutwyler[3], Adrien Michel[3], Jonathan Koh[4]**
[1] Faculty of Geosciences and Environment, University of Lausanne, Switzerland
[2] Expertise Center for Climate Extremes (ECCE), Faculty of Business and Economics (HEC) - Faculty of Geosciences and Environment,  University of Lausanne, Switzerland
[3] Federal Office of Meteorology and Climatology MeteoSwiss, Switzerland
[4] Oeschger Centre for Climate Change Research, University of Bern, Switzerland

**Recommendation 1:** *Develop Hybrid AI-Physical Models*: Emphasize the integration of AI and physical modeling for improved reliability, especially for longer prediction horizons, acknowledging the delicate balance between knowledge-based and data-driven components required for optimal performance.

**Recommendation 2:** *Emphasize Robustness in AI Downscaling Approaches*, favoring techniques that respect physical laws, preserve inter-variable dependencies and spatial structures, and accurately represent extremes at the local scale.

**Recommendation 3:** *Promote Inclusive Model Development*: Ensure Earth System Model development is open and accessible to diverse stakeholders, enabling forecasters, the public, and AI/statistics experts to use, develop, and engage with the model and its predictions/projections.

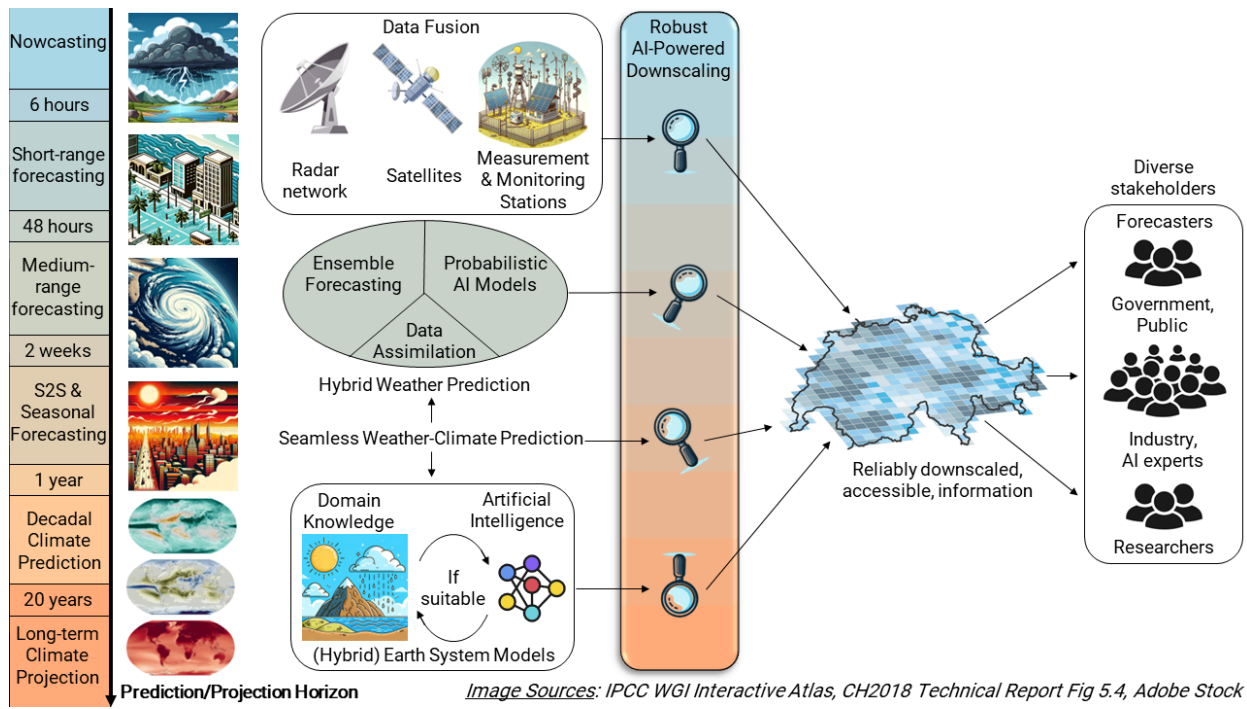

**Figure Caption:** Advancements in data collection, data access, hybrid AI-physical Earth system modeling, and downscaling empower stakeholders with increased accessibility to local predictions and projections, encouraging collaborative efforts across disciplines to improve climate change preparedness.

Here, we review how machine learning has transformed our ability to model the Earth system, and how we expect recent breakthroughs to benefit end-users in Switzerland in the near future.

### 5.2.1 Limitations of traditional Earth system models

Earth system models (ESMs), which encode our knowledge of the Earth system into equations that can be integrated forward in time, serve the double purpose of understanding and prediction (Prinn 2013, Held 2005). Weather forecasting predicts the atmospheric state at short timescales, typically less than two weeks. At longer timescales, chaos makes the exact atmospheric state unpredictable and ESMs are used to predict or project climate statistics. Except for seamless weather-climate prediction setups (Hoskins, 2013), the ESMs used for numerical weather prediction (NWP) and climate projections are different and face their own sets of challenges.

While significant progress has been achieved in data assimilation (Courtier et al., 1994; de Rosnay et al., 2022), physical process representation (Bauer et al., 2015), and ensemble forecasting including extreme events (BenBouallègue et al., 2019), the computational cost of NWP remains a major bottleneck, forcing a delicate balance between ensemble size and model resolution. Progress on leveraging the capabilities of current and emerging supercomputer architectures for ESMs and NWP is slower than expected (Schulthess et al., 2018). High-fidelity, high-resolution ESMs, will thus require very substantial investments in high-performance supercomputers (Bauer et al., 2021) and re-engineering the model codes (Tal Ben-Nun et al., 2022).

On long timescales, ESMs used for climate projections face several challenges in faithfully representing Earth system processes in the atmosphere, ocean, and land, even if they accurately simulate the rise in global mean temperatures linked to anthropogenic forcing (Eyring et al., 2021). In the atmosphere, uncertainties in climate projections primarily arise from convection and cloud-aerosol interactions (Rosenfeld et al., 2014). These lead to structural biases in warming patterns (Andrews et al., 2022) and precipitation (Palmer and Stevens, 2019). Global storm-resolving models can mitigate some of these biases by explicitly simulating deep convection (Stevens et al., 2020), but cannot be easily calibrated and used for century-long projections because of their high computational cost; these models also still rely on low-level clouds, aerosols, and microphysics parameterizations (Schneider et al., 2017). In the ocean, uncertainties persist due to unresolved mesoscale eddies and turbulent processes (Couldrey et al., 2021). Predicting the land carbon sink remains a challenge (Friedlingstein et al., 2022), progressively addressed through observationally-driven reductions in model uncertainties (Dagon et al., 2020), notably those related to extremes (Reichstein et al., 2013), and land water and carbon cycles (Gentine et al., 2019).

Therefore, the ESMs' role in understanding the climate system remains paramount in part because they allow us to explore counterfactuals that are relatively unconstrained by observations, such as a variety of forcings and the long-term climate response to forcings (Balaji et al., 2022). However, current roadblocks of traditional ESMs, such as providing consistent bounds on equilibrium climate sensitivity (Forster et al., 2021), appear inconsistent with our quickly increasing computational and observational capabilities.

### 5.2.2 Can we use unconstrained AI for weather and climate prediction?

"Pure", or "hard" AI (Chantry et al., 2021), unconstrained by the structure of dynamical equations, can leverage the "deluge" of Earth system data, from remote sensing to in-situ measurements via citizen observations (Reichstein et al., 2019).

This progress has already revolutionized NWP (BenBouallègue et al., 2023). In nowcasting, where forecasts are made for the next (0–3) hours, deep generative models used for video prediction (Ravuri et al., 2021) are particularly suitable for combining different observational sources and making fast inferences. Medium-range (up to a couple of weeks) forecasting has recently witnessed high-performance, high-resolution, purely data-driven forecasts, notably GraphCast (Lam et al., 2022), Pangu-Weather (Bi et al., 2023), and FourCastNet (Pathak et al., 2022; Bonev et al., 2023). Their deterministic error is on par with and in some cases even lower than that of our best NWP models (Rasp et al., 2023), despite making inferences using one GPU instead of thousands of CPUs. As generative modeling allows the creation of "infinite ensembles", we expect rapid progress in "pure AI" probabilistic forecasts following successful prototypes for, e.g., atmospheric river prediction (Chapman et al., 2022) and medium-range forecasting (Garg et al., 2022; Price et al., 2023).

In climate change research, flexible machine learning (ML) approaches have advanced our understanding of connected, nonlinear, or non-Gaussian Earth system processes (Huntingford et al., 2019). ML can offer multiple benefits, including accelerated computing, enhanced data fidelity, and added value to climate simulation outputs.

For acceleration, ML models are trained to closely mimic Earth process models, leveraging the inherent speed of ML algorithms, including neural networks, which typically involve fewer floating-point operations and benefit from GPU computing. This approach has a history of improving the temporal resolutions of radiative transfer models in weather (Chevallier et al., 1998) and climate (Pal et al., 2019) simulations. It can also expedite the execution of computationally expensive parameterizations (modules) in ESMs, such as convection (O'Gorman and Dwyer, 2018) and complex bin microphysical schemes (Gettelman et al., 2021).

In terms of data fidelity, ML enhances our Earth system representation by learning from high-fidelity data, such as observations, reanalyses, or high-resolution model runs, such as "digital twins" (Bauer et al., 2021). These "digital twins" are too costly for extended simulations but offer millions of samples for developing ML parameterizations. Recent applications include enhancing subgrid-scale parameterization in climate models by learning from storm-resolving simulations, which improves various critical climate model statistics, from extreme precipitation distributions to tropical wave spectra (Rasp et al., 2018). Deep learning and symbolic regression have also been employed for subgrid parameterization (Ross et al., 2023) and vertical mixing (Sane et al., 2023) in the ocean.

ML has also demonstrated its value in augmenting observational datasets, as shown by its use in estimating spatiotemporal land-water fluxes through random forests (Jung et al., 2011) and ocean carbon fluxes via neural networks (Landschützer et al., 2013). Recent applications, like "ClimateNet", showcase its role in automating climate analysis by segmenting extreme weather events in climate model output (Prabhat et al., 2021). Generative models such as Gaussian processes (Camps-Valls et al., 2016) can also add value by inherently estimating uncertainty, with applications to, e.g., constraining sensitivities from climate model ensembles (Watson-Parris et al., 2020). Finally, foundation models (Bommasani et al., 2021), have emerged as promising tools for weather and climate applications, as exemplified by ClimaX (Nguyen et al., 2023). These large neural networks, originally pre-trained on diverse data, can be fine-tuned with minimal samples for specific tasks like forecasting and downscaling.

The limitations of "pure AI" are most visible with sparse or difficult-to-process data, such as ground-based observations (Schultz et al., 2021). Standardized benchmark datasets are being developed for weather and climate applications to address this issue (Dueben et al., 2023). Even in cases where data are plentiful, purely data-driven weather forecasts often rely on meteorological reanalysis data for success, indirectly tying their quality to traditional ESMs and data assimilation systems. Challenges in climate modeling and projection include poor out-of-climate generalization (O'Gorman and Dwyer., 2018), instabilities from AI components-ESM interactions (Brenowitz et al., 2018), disparities in offline and online skill (Brenowitz et al., 2020), and physical inconsistencies, such as the violation of conservation laws (Beucler et al., 2019).

Many of the aforementioned challenges can be resolved by combining data-driven models for accurate in-sample prediction and knowledge-based models for induction. This is broadly referred to as knowledge-guided ML (Karpatne et al., 2022), which includes physics-guided ML (Willard et al., 2020) and spans methods ranging from physics-informed neural networks (Raissi et al., 2019) to Gaussian processes-based calibration of physical model parameters (Cleary et al., 2021).

### 5.2.3 Hybrid Climate-AI Modeling: Towards the best of both worlds

Finding the optimal balance between knowledge-based and data-driven components is delicate; Chantry et al. (2021) argue that the longer the prediction timescale, the softer the AI should be. This seems broadly consistent with the recent breakthroughs in deep learning applications for NWP, starting with nowcasting (Sonderby et al., 2020; Leinonen et al., 2023) and progressively permeating medium-range forecasting.

Irrgang et al. (2021) coined the term "neural ESMs" to describe hybrid models that can reproduce and predict out-of-distribution samples and extreme events, perform constrained simulations that obey physical conservation laws, include measures to self-validate and self-correct, and allow replicability and interpretability.

Practical progress towards this framework has been made in recent years, such as architecture-based constraints to ensure conservation laws (Beucler et al., 2021), incorporation of symmetry to enhance generalization (Wang et al., 2022), coupled online learning to mitigate instabilities and biases (Lopez-Gomez et al., 2022), input restrictions to improve stability (Bretherton et al., 2022), causal model evaluation (Nowack et al., 2020) and causal deep learning (Iglesias-Suarez et al., 2023) to respect underlying physical processes, data-driven equation discovery (Grundner et al., 2023), and the use of transfer learning and climate-invariant inputs to enhance generalization (Chattopadhyay et al., 2020; Beucler et al., 2021). Recently, NeuralGCM has emerged as the first fully differentiable hybrid general circulation model, coupling a dynamical core that spectrally solves the primitive equations with neural networks trained to accurately represent physical processes for optimal short-term weather prediction (Kochkov et al., 2023).

Regional and global climate, pure AI or hybrid climate-AI models frequently exhibit grid spacings exceeding 10 kilometers, making them of limited usefulness for risk assessment. Impact models indeed typically require climate inputs at a local scale of, at most a few kilometers. This gap can be filled to some extent by downscaling algorithms.

### 5.2.4 From climate data to actionable information: AI-assisted downscaling

AI, and in particular super-resolution (Wang et al., 2020), has already proven its potential to revolutionize statistical downscaling, which predicts societally relevant variables at the local scale from coarser-scale ESM output. We briefly survey some of the recent developments in this field.

Convolutional neural networks (CNNs) have been successfully used to super-resolve precipitation projections from ESMs (Vandal et al., 2019; Rampal et al., 2022), satellite images (Pouliot et al., 2018), radar reflectivity scans (Geiss and Hardin, 2020), winds over complex terrain (Dujardin and Lehning, 2022), and idealized turbulent flows (Fukami et al., 2019) with high pixel-wise accuracy.

Despite outperforming conventional techniques such as bicubic interpolation (as shown in Baño-Medina et al, 2020) and offering the possibility of increasing temporal resolution by using multiple channels (Serifi et al, 2021) or a recurrent structure (Harilal et al, 2021), CNNs trained with standard loss functions typically underestimate extremes (Sachindra et al, 2018) as they tend to predict the average of all possible solutions to minimize the error at each pixel.

While Ghosh et al. (2008) and Sachindra et al. (2018) demonstrated that relevance vector machines (a Bayesian approach to learning probabilistic sparse generalized linear models, Tipping et al., 1999) enhance the downscaling of river streamflow and precipitation extremes, recent investigations have shifted towards generative modeling, specifically focusing on conditional Generative Adversarial Networks (cGANs). For example, Stengel et al. (2020) used a sequence of two cGANs to super-resolve wind velocity and solar irradiance on the 2km scale using 100 km-scale climate model output. They showed that the resulting spatial variability (as quantified by the turbulent kinetic energy spectra and solar irradiance semivariograms) was more consistent with that of high-resolution climate model outputs than the one generated by interpolation and simple CNNs, consistent with studies using GANs to downscale precipitation (Watson et al., 2020). Groenke et al. (2020) adapted recent work in normalizing flows (Rezende et al., 2015) for variational inference (Grover et al., 2020) to develop an unsupervised neural network approach that generates the joint distribution of high- and low-resolution climate maps. Leinonen et al. (2020) used the stochastic nature of GANs to generate ensembles of time-evolving, super-resolved radar-measured precipitation over Switzerland, and satellite-derived cloud optical thickness fields.

Miralles et al., (2022) used a similar Wasserstein recurrent GAN architecture (Gulrajani et al., 2017) to downscale meteorological reanalysis winds to storm-resolution over Switzerland. In 2023, diffusion models witnessed their first atmospheric applications (Leinonen et al., 2023), including solar irradiance (Hatanaka et al., 2023) and precipitation (Addison et al., 2022) downscaling. Notably, Mardani et al., (2023) simultaneously downscaled winds, temperature, and radar reflectivity using a two-step approach akin to that from Price et al. (2022).

For downscaled (or bias-adjusted, or more broadly post-processed) outputs to be actionable, they must ideally meet several key requirements, including:
- several features (e.g., change signal for the mean and extremes) should be conserved;
- inter-variable dependencies should be well preserved;
- extremes should be reliably represented at a very local scale;
- misrepresented relations in the original model output should be corrected;
- spatial consistency of the fields should be ensured;
- the temporal resolution should be sub-daily.

To our knowledge, there exists no universal post-processing/downscaling approach fulfilling all these requirements at the same time. The available methods typically satisfy some of them only and, in practice, the employed technique is chosen depending on the specific application. Further research is therefore needed to fill these gaps.

### 5.2.5 Recommendations with a focus on Switzerland

There is high confidence that climate change in Switzerland will increase temperatures and extreme summer rainfall, change the seasonality of river discharge, and affect groundwater, water temperatures, snow, and glaciers. Adapting to these changes requires accessible information at the local scale. The intricate topography of Switzerland, like the Jura Mountains and the Alps, poses a tremendous challenge in generating climate projections at fine scales. In pursuit of this goal, the Swiss climate modeling community has made notable strides in recent years, advancing the precision of climate-time scale simulations to encompass the Alpine region and Europe at model resolutions of mere kilometers (Ban et al., 2014; Ban et al., 2020; Ban et al., 2021; Leutwyler et al., 2017; Schär et al. 2020). These achievements will enable kilometer-resolution regional climate projections and bulk assessment of changes for an entire region, like the Swiss Plateau (Langhans et al., 2012).

However, the effective modeling and evaluation of climate impacts on these diverse terrains necessitate even more granular, higher resolution, and bias-free data of multiple meteorological variables. For example, creating a dynamic model of the Morteratsch Glacier requires incredibly detailed boundary conditions that include data on precipitation, temperature, and the surface radiation balance, potentially down to hectometer or even decameter resolutions. Achieving such fine granularity in the foreseeable future will likely depend on downscaling techniques that can bridge the gap between kilometer-scale simulations and the microclimates of these landscapes.

In the CH2018 Swiss climate change scenarios (Fischer et al., 2022), the quantile mapping (QM) applied for downscaling and bias adjustment has been performed at the univariate level (variable by variable; Feigenwinter et al. 2018), which implies that inter-variable dependence might have been violated, especially for extremes (Michel et al., 2021). Moreover, biases in the inter-variable dependencies inherent to the raw outputs from GCMs and RCMs cannot be corrected by univariate QM. Reliably representing the extremal inter-variable dependence is essential to forecast/project compound events and for impact models (e.g., consistent 2m temperature, precipitation, and surface radiation balance for glacier models). In the upcoming CH2025 climate change scenarios, inter-variable dependence should be partially preserved owing to multivariate bias-correction methods (e.g., Ortner et al., 2023; Cannon et al., 2018), but new techniques will be needed to ensure spatial consistency of the fields, as this is essential for impact assessment of short and heavy rainfall.

Efforts to find out the best synergies between physical modeling, AI methods, and statistical models grounded in probability theory for forecasting/projecting extremes (i.e., to develop "physically and statistically-informed ML"), such as those carried out by the recently created Expertise Center for Climate Extremes (ECCE) at the University of Lausanne, are essential to moving a step forward on these methodological aspects. Extreme-value theory (EVT), a field that has sustained a long tradition in Switzerland, for example in insurance (Embrechts et al. 1997) or environmental applications (Davison et al. 2012), provides such an example of an elegant and mathematically justified framework to estimate the probabilities of rare events by extrapolating beyond the range of the data. Its combination with machine learning tools is recent, and studies have mainly focused on adapting the loss functions used in ML algorithms to focus predominantly on extremes and to incorporate the modulation of these events with relevant predictors. This has been done to predict conditional tail distributions using gradient boosting (Koh, 2023), random forest (Gnecco et al., 2023), and neural network architectures (Richards et al., 2022, Pasche and Engelke, 2023). The paucity of data in the distributional tail makes it challenging to fit well and verify these models, highlighting the need for further research. Another central task then consists of elaborating detailed information and scenarios about extremes at a very local scale, combining them with vulnerability models and exposure data to provide risk assessments, and informing the cantons, the confederation, and other stakeholders (such as insurance companies).

Finally, data-driven modeling has a longstanding history in Swiss NWP applications, notably for mesocyclone and severe thunderstorm tracking based on radar observations (Hering et al., 2004) and satellite images. With the advent of deep learning, the use of AI has accelerated. Neural networks now play a pivotal role in diverse tasks, such as post-processing cloud cover ensemble forecasts (Dai and Hemri, 2021) and real-time nowcasting of hail and lightning events (Leinonen et al., 2022). As NWP practitioners acclimate to AI tools, a prevailing consensus emerges, favoring their supplementary role alongside existing methodologies rather than outright replacement (Boukarba et al., 2020). This consensus emphasizes the enduring importance of domain knowledge, fostering the development of hybrid models wherein human expertise is seamlessly integrated into AI frameworks for weather applications (e.g., Zanetta et al., 2023) and beyond, a trend we anticipate will persist in the foreseeable future.

# Bibliography

Addison, H., Kendon, E., Ravuri, S., Aitchison, L., & Watson, P. A. (2022). Machine learning emulation of a local-scale UK climate model. arXiv preprint arXiv:2211.16116.

Andrews, T., Bodas-Salcedo, A., Gregory, J. M., Dong, Y., Armour, K. C., Paynter, D., ... & Liu, C. (2022). On the effect of historical SST patterns on radiative feedback. *Journal of Geophysical Research: Atmospheres*, *127*(18), e2022JD036675.

Balaji, V., Couvreux, F., Deshayes, J., Gautrais, J., Hourdin, F., & Rio, C. (2022). Are general circulation models obsolete?. *Proceedings of the National Academy of Sciences*, *119*(47), e2202075119.

Ban, Nikolina, Juerg Schmidli, and Christoph Schär. "Evaluation of the convection-resolving regional climate modeling approach in decade-long simulations." *Journal of Geophysical Research: Atmospheres* 119.13 (2014): 7889-7907.

Ban, N., Rajczak, J., Schmidli, J., & Schär, C. (2020). Analysis of Alpine precipitation extremes using generalized extreme value theory in convection-resolving climate simulations. *Climate Dynamics*, *55*(1-2), 61-75.

Ban, N., Caillaud, C., Coppola, E., Pichelli, E., Sobolowski, S., Adinolfi, M., ... & Zander, M. J. (2021). The first multi-model ensemble of regional climate simulations at kilometer-scale resolution, part I: evaluation of precipitation. *Climate Dynamics*, *57*, 275-302.

Baño-Medina, J., Manzanas, R., & Gutiérrez, J. M. (2020). Configuration and intercomparison of deep learning neural models for statistical downscaling. *Geoscientific Model Development*, *13*(4), 2109-2124.

Bauer, P., Thorpe, A., & Brunet, G. (2015). The quiet revolution of numerical weather prediction. *Nature*, *525*(7567), 47-55.

Bauer, P., Dueben, P. D., Hoefler, T., Quintino, T., Schulthess, T. C., & Wedi, N. P. (2021). The digital revolution of Earth-system science. *Nature Computational Science*, *1*(2), 104-113.

Bauer, P., Stevens, B., & Hazeleger, W. (2021). A digital twin of Earth for the green transition. *Nature Climate Change*, *11*(2), 80-83.

Ben Bouallègue, Z., Magnusson, L., Haiden, T., & Richardson, D. S. (2019). Monitoring trends in ensemble forecast performance focusing on surface variables and high-impact events. *Quarterly Journal of the Royal Meteorological Society*, *145*(721), 1741-1755.

Ben-Bouallegue, Z., Clare, M. C., Magnusson, L., Gascon, E., Maier-Gerber, M., Janousek, M., ... & Pappenberger, F. (2023). The rise of data-driven weather forecasting. *arXiv preprint arXiv:2307.10128*.

Ben-Nun, T., Groner, L., Deconinck, F., Wicky, T., Davis, E., Dahm, J., ... & Hoefler, T. (2022, November). Productive performance engineering for weather and climate modeling with python. In *SC22: International Conference for High Performance Computing, Networking, Storage and Analysis* (pp. 1-14). IEEE.

Beucler, T., Rasp, S., Pritchard, M., & Gentine, P. (2019). Achieving conservation of energy in neural network emulators for climate modeling. *arXiv preprint arXiv:1906.06622*.

Beucler, T., Pritchard, M., Rasp, S., Ott, J., Baldi, P., & Gentine, P. (2021). Enforcing analytic constraints in neural networks emulating physical systems. *Physical Review Letters*, *126*(9), 098302.

Beucler, T., Pritchard, M., Yuval, J., Gupta, A., Peng, L., Rasp, S., ... & Gentine, P. (2021). Climate-invariant machine learning. *arXiv preprint arXiv:2112.08440*.

Bi, K., Xie, L., Zhang, H., Chen, X., Gu, X., & Tian, Q. (2023). Accurate medium-range global weather forecasting with 3D neural networks. *Nature*, *619*(7970), 533-538.

Bommasani, R., Hudson, D. A., Adeli, E., Altman, R., Arora, S., von Arx, S., ... & Liang, P. (2021). On the opportunities and risks of foundation models. *arXiv preprint arXiv:2108.07258*.

Bonev, B., Kurth, T., Hundt, C., Pathak, J., Baust, M., Kashinath, K., & Anandkumar, A. (2023). Spherical Fourier Neural Operators: Learning Stable Dynamics on the Sphere. *arXiv preprint arXiv:2306.03838*.

Boukabara, S. A., Krasnopolsky, V., Penny, S. G., Stewart, J. Q., McGovern, A., Hall, D., ... & Hoffman, R. N. (2020). Outlook for exploiting artificial intelligence in the earth and environmental sciences. *Bulletin of the American Meteorological Society*, 1-53.

Brenowitz, N. D., & Bretherton, C. S. (2018). Prognostic validation of a neural network unified physics parameterization. *Geophysical Research Letters*, *45*(12), 6289-6298.

Brenowitz, N. D., Beucler, T., Pritchard, M., & Bretherton, C. S. (2020). Interpreting and stabilizing machine-learning parametrizations of convection. *Journal of the Atmospheric Sciences*, *77*(12), 4357-4375.

Bretherton, C. S., Henn, B., Kwa, A., Brenowitz, N. D., Watt-Meyer, O., McGibbon, J., ... & Harris, L. (2022). Correcting coarse-grid weather and climate models by machine learning from global storm-resolving simulations. *Journal of Advances in Modeling Earth Systems*, *14*(2), e2021MS002794.

Camps-Valls, G., Verrelst, J., Munoz-Mari, J., Laparra, V., Mateo-Jimenez, F., & Gomez-Dans, J. (2016). A survey on Gaussian processes for earth-observation data analysis: A comprehensive investigation. *IEEE Geoscience and Remote Sensing Magazine*, *4*(2), 58-78.

Cannon, A. J. (2018). Multivariate quantile mapping bias correction: an N-dimensional probability density function transform for climate model simulations of multiple variables. *Climate dynamics*, *50*, 31-49.

Chantry, M., Christensen, H., Dueben, P., & Palmer, T. (2021). Opportunities and challenges for machine learning in weather and climate modelling: hard, medium and soft AI. *Philosophical Transactions of the Royal Society A*, *379*(2194), 20200083.

Chapman, W. E., Subramanian, A. C., Delle Monache, L., Xie, S. P., & Ralph, F. M. (2019). Improving atmospheric river forecasts with machine learning. *Geophysical Research Letters*, *46*(17-18), 10627-10635.

Chattopadhyay, A., Subel, A., & Hassanzadeh, P. (2020). Data-driven super-parameterization using deep learning: Experimentation with multiscale Lorenz 96 systems and transfer learning. *Journal of Advances in Modeling Earth Systems*, *12*(11), e2020MS002084.

Chevallier, F., Chéruy, F., Scott, N. A., & Chédin, A. (1998). A neural network approach for a fast and accurate computation of a longwave radiative budget. *Journal of applied meteorology*, *37*(11), 1385-1397.

Cleary, E., Garbuno-Inigo, A., Lan, S., Schneider, T., & Stuart, A. M. (2021). Calibrate, emulate, sample. *Journal of Computational Physics*, *424*, 109716.

Couldrey, M. P., Gregory, J. M., Boeira Dias, F., Dobrohotoff, P., Domingues, C. M., Garuba, O., ... & Zanna, L. (2021). What causes the spread of model projections of ocean dynamic sea-level change in response to greenhouse gas forcing?. *Climate Dynamics*, *56*(1-2), 155-187.

Courtier, P., Thépaut, J. N., & Hollingsworth, A. (1994). A strategy for operational implementation of 4D-Var, using an incremental approach. *Quarterly Journal of the Royal Meteorological Society*, *120*(519), 1367-1387.

Dagon, K., Sanderson, B. M., Fisher, R. A., & Lawrence, D. M. (2020). A machine learning approach to emulation and biophysical parameter estimation with the Community Land Model, version 5. *Advances in Statistical Climatology, Meteorology and Oceanography*, *6*(2), 223-244.

Dai, Y. & Hemri, S. (2021). Spatially Coherent Postprocessing of Cloud Cover Ensemble Forecasts.
*Monthly Weather Review, 149*(12), 3923–3937.

Davison, A. C., Padoan, S. A., & Ribatet, M. (2012). Statistical modeling of spatial extremes. Statistical Science 27(2), 161–186.

de Rosnay, P., Browne, P., de Boisséson, E., Fairbairn, D., Hirahara, Y., Ochi, K., ... & Rabier, F. (2022). Coupled data assimilation at ECMWF: Current status, challenges and future developments. Quarterly Journal of the Royal Meteorological Society, 148(747), 2672-2702.

Dueben, P. D., Schultz, M. G., Chantry, M., Gagne, D. J., Hall, D. M., & McGovern, A. (2022). Challenges and benchmark datasets for machine learning in the atmospheric sciences: Definition, status, and outlook. *Artificial Intelligence for the Earth Systems*, *1*(3), e210002.

Dujardin, J., & Lehning, M. (2022). Wind-Topo: Downscaling near-surface wind fields to high-resolution topography in highly complex terrain with deep learning. Quarterly Journal of the Royal Meteorological Society, 148(744), 1368-1388.

Embrechts, P., Klüppelberg, C., & Mikosch, T. (1997). Modelling extremal events: for insurance and finance. Springer Berlin, Heidelberg.

Eyring, V., Gillett, N. P., Achuta Rao, K. M., Barimalala, R., Barreiro Parrillo, M., Bellouin, N., ... & Sun, Y. (2021). Human Influence on the Climate System (Chapter 3).

Feigenwinter, I., Kotlarski, S., Casanueva, A., Schwierz, C., & Liniger, M. A. (2018). *Exploring quantile mapping as a tool to produce user-tailored climate scenarios for Switzerland*. MeteoSchweiz.

Fischer, A. M., Strassmann, K. M., Croci-Maspoli, M., Hama, A. M., Knutti, R., Kotlarski, S., ... & Zubler, E. M. (2022). Climate scenarios for Switzerland CH2018–approach and implications. Climate services, 26, 100288.

Friedlingstein, P., Jones, M. W., O'sullivan, M., Andrew, R. M., Bakker, D. C., Hauck, J., ... & Zeng, J. (2022). Global carbon budget 2021. *Earth System Science Data*, *14*(4), 1917-2005.

Forster, P., Storelvmo, T., Armour, K., Collins, W., Dufresne, J. L., Frame, D., ... & Zhang, H. (2021). The Earth's energy budget, climate feedbacks, and climate sensitivity.

Fukami, K., Fukagata, K., & Taira, K. (2019). Super-resolution reconstruction of turbulent flows with machine learning. *Journal of Fluid Mechanics*, *870*, 106-120.

Garg, S., Rasp, S., & Thuerey, N. (2022). WeatherBench Probability: A benchmark dataset for probabilistic medium-range weather forecasting along with deep learning baseline models. *arXiv preprint arXiv:2205.00865*.

Geiss, A., & Hardin, J. C. (2020). Radar super resolution using a deep convolutional neural network. Journal of Atmospheric and Oceanic Technology, 37(12), 2197-2207.

Gentine, P., Green, J. K., Guérin, M., Humphrey, V., Seneviratne, S. I., Zhang, Y., & Zhou, S. (2019). Coupling between the terrestrial carbon and water cycles—a review. *Environmental Research Letters*, *14*(8), 083003.

Gettelman, A., Gagne, D. J., Chen, C. C., Christensen, M. W., Lebo, Z. J., Morrison, H., & Gantos, G. (2021). Machine learning the warm rain process. *Journal of Advances in Modeling Earth Systems*, *13*(2), e2020MS002268.

Ghosh, S., & Mujumdar, P. P. (2008). Statistical downscaling of GCM simulations to streamflow using relevance vector machine. Advances in water resources, 31(1), 132-146.

Gnecco, N., Terefe, E. M., & Engelke, S. (2023). Extremal random forest. *arXiv preprint arXiv:2201.12865*

Groenke, B., Madaus, L., & Monteleoni, C. (2020, September). ClimAlign: Unsupervised statistical downscaling of climate variables via normalizing flows. In Proceedings of the 10th International Conference on Climate Informatics (pp. 60-66).

Grover, A., Chute, C., Shu, R., Cao, Z., & Ermon, S. (2020, April). Alignflow: Cycle consistent learning from multiple domains via normalizing flows. In Proceedings of the AAAI Conference on Artificial Intelligence (Vol. 34, No. 04, pp. 4028-4035).

Grundner, A., Beucler, T., Gentine, P., & Eyring, V. (2023). Data-Driven Equation Discovery of a Cloud Cover Parameterization. *arXiv preprint arXiv:2304.08063*.

Gulrajani, I., Ahmed, F., Arjovsky, M., Dumoulin, V., & Courville, A. C. (2017). Improved training of wasserstein gans. Advances in neural information processing systems, 30.

Hatanaka, Y., Glaser, Y., Galgon, G., Torri, G., & Sadowski, P. (2023). Diffusion Models for High-Resolution Solar Forecasts. arXiv preprint arXiv:2302.00170.

Harilal, N., Singh, M., & Bhatia, U. (2021). Augmented convolutional LSTMs for generation of high-resolution climate change projections. IEEE Access, 9, 25208-25218.

Held, I. M. (2005). The gap between simulation and understanding in climate modeling. *Bulletin of the American Meteorological Society*, *86*(11), 1609-1614.

Hering, A. M., Morel, C., Galli, G., Sénési, S., Ambrosetti, P., &  Boscacci, M (2004). Nowcasting thunderstorms in the Alpine region using a radar based adaptive thresholding scheme. *Proceedings of the 3th European Conference on Radar in Meteorology and Hydrology, Gotland, Sweden, 6–10 September 2004; Copernicus: Gottingen, Germany, 2004*, 206-211.

Hoskins, B. (2013). The potential for skill across the range of the seamless weather-climate prediction problem: a stimulus for our science. *Quarterly Journal of the Royal Meteorological Society*, *139*(672), 573-584.

Huntingford, C., Jeffers, E. S., Bonsall, M. B., Christensen, H. M., Lees, T., & Yang, H. (2019). Machine learning and artificial intelligence to aid climate change research and preparedness. *Environmental Research Letters*, *14*(12), 124007.

Iglesias-Suarez, F., Gentine, P., Solino-Fernandez, B., Beucler, T., Pritchard, M., Runge, J., & Eyring, V. (2023). Causally-informed deep learning to improve climate models and projections. *arXiv preprint arXiv:2304.12952*.

Irrgang, C., Boers, N., Sonnewald, M., Barnes, E. A., Kadow, C., Staneva, J., & Saynisch-Wagner, J. (2021). Towards neural Earth system modelling by integrating artificial intelligence in Earth system science. *Nature Machine Intelligence*, *3*(8), 667-674.

Jung, M., Reichstein, M., Margolis, H. A., Cescatti, A., Richardson, A. D., Arain, M. A., ... & Williams, C. (2011). Global patterns of land-atmosphere fluxes of carbon dioxide, latent heat, and sensible heat derived from eddy covariance, satellite, and meteorological observations. *Journal of Geophysical Research: Biogeosciences*, *116*(G3).

Karpatne, A., Kannan, R., & Kumar, V. (Eds.). (2022). *Knowledge Guided Machine Learning: Accelerating Discovery Using Scientific Knowledge and Data*. CRC Press.

Kochkov, D., Yuval, J., Langmore, I., Norgaard, P., Smith, J., Mooers, G., ... & Hoyer, S. (2023). Neural General Circulation Models. *arXiv preprint arXiv:2311.07222*.

Koh, J. (2023). Gradient boosting with extreme-value theory for wildfire prediction. *Extremes 26, 273–299*

Langhans, W., Schmidli, J., & Schär, C. (2012). Bulk convergence of cloud-resolving simulations of moist convection over complex terrain. *Journal of the Atmospheric Sciences*, *69*(7), 2207-2228.

Lam, R., Sanchez-Gonzalez, A., Willson, M., Wirnsberger, P., Fortunato, M., Pritzel, A., ... & Battaglia, P. (2023). GraphCast: Learning skillful medium-range global weather forecasting. *Science 382, 1416-1421*.

Landschützer, P., Gruber, N., Bakker, D. C., Schuster, U., Nakaoka, S. I., Payne, M. R., ... & Zeng, J. (2013). A neural network-based estimate of the seasonal to inter-annual variability of the Atlantic Ocean carbon sink. *Biogeosciences*, *10*(11), 7793-7815.

Leinonen, J., Nerini, D., & Berne, A. (2020). Stochastic super-resolution for downscaling time-evolving atmospheric fields with a generative adversarial network. IEEE Transactions on Geoscience and Remote Sensing, 59(9), 7211-7223.

Leinonen, J., Hamann, U., and Germann, U. Seamless Lightning Nowcasting with Recurrent-Convolutional Deep Learning (2022). *Artificial Intelligence for the Earth Systems, 1*(4), e220043

Leinonen, J., Hamann, U., Sideris, I. V., & Germann, U. (2023). Thunderstorm Nowcasting With Deep Learning: A Multi-Hazard Data Fusion Model. *Geophysical Research Letters*, *50*(8), e2022GL101626.

Leinonen, J., Hamann, U., Nerini, D., Germann, U., & Franch, G. (2023). Latent diffusion models for generative precipitation nowcasting with accurate uncertainty quantification. arXiv preprint arXiv:2304.12891.

Leutwyler, D., Lüthi, D., Ban, N., Fuhrer, O., & Schär, C. (2017). Evaluation of the convection-resolving climate modeling approach on continental scales. *Journal of Geophysical Research: Atmospheres*, *122*(10), 5237-5258.

Lopez-Gomez, I., Christopoulos, C., Langeland Ervik, H. L., Dunbar, O. R., Cohen, Y., & Schneider, T. (2022). Training physics-based machine-learning parameterizations with gradient-free ensemble Kalman methods. *Journal of Advances in Modeling Earth Systems*, *14*(8), e2022MS003105.

Mardani, M., Brenowitz, N., Cohen, Y., Pathak, J., Chen, C. Y., Liu, C. C., ... & Pritchard, M. (2023). Generative Residual Diffusion Modeling for Km-scale Atmospheric Downscaling. arXiv preprint arXiv:2309.15214.

Michel, A., Sharma, V., Lehning, M., & Huwald, H. (2021). Climate change scenarios at hourly time-step over Switzerland from an enhanced temporal downscaling approach. *International Journal of Climatology*, *41*(6), 3503-3522.

Miralles, O., Steinfeld, D., Martius, O., & Davison, A. C. (2022). Downscaling of Historical Wind Fields over Switzerland Using Generative Adversarial Networks. Artificial Intelligence for the Earth Systems, 1(4), e220018.

Nguyen, T., Brandstetter, J., Kapoor, A., Gupta, J. K., & Grover, A. (2023). ClimaX: A foundation model for weather and climate. *arXiv preprint arXiv:2301.10343*.

Nowack, P., Runge, J., Eyring, V., & Haigh, J. D. (2020). Causal networks for climate model evaluation and constrained projections. Nature communications, 11(1), 1415.

O'Gorman, P. A., & Dwyer, J. G. (2018). Using machine learning to parameterize moist convection: Potential for modeling of climate, climate change, and extreme events. *Journal of Advances in Modeling Earth Systems*, *10*(10), 2548-2563.

Ortner, G., Michel, A., Spieler, M., Christen, M., Bühler, Y., Bründl, M., & Bresch, D. N. (2023). Assessing climate change impacts on snow avalanche hazard. *Available at SSRN 4530305*.

Pal, A., Mahajan, S., & Norman, M. R. (2019). Using deep neural networks as cost-effective surrogate models for super-parameterized E3SM radiative transfer. *Geophysical Research Letters*, *46*(11), 6069-6079.

Palmer, T., & Stevens, B. (2019). The scientific challenge of understanding and estimating climate change. *Proceedings of the National Academy of Sciences*, *116*(49), 24390-24395.

Pasche, O. C., & Engelke, S. (2023). Neural networks for extreme quantile regression with an application to forecasting of flood risk. *arXiv preprint arXiv:2208.07590*

Pathak, J., Subramanian, S., Harrington, P., Raja, S., Chattopadhyay, A., Mardani, M., ... & Anandkumar, A. (2022). Fourcastnet: A global data-driven high-resolution weather model using adaptive fourier neural operators. *arXiv preprint arXiv:2202.11214*.

Pouliot, D., Latifovic, R., Pasher, J., & Duffe, J. (2018). Landsat super-resolution enhancement using convolution neural networks and Sentinel-2 for training. *Remote Sensing*, *10*(3), 394.

Prabhat, Kashinath, K., Mudigonda, M., Kim, S., Kapp-Schwoerer, L., Graubner, A., Karaismailoglu, E., ... & Collins, W. (2021). ClimateNet: An expert-labeled open dataset and deep learning architecture for enabling high-precision analyses of extreme weather. *Geoscientific Model Development*, *14*(1), 107-124.

Price, I., & Rasp, S. (2022, May). Increasing the accuracy and resolution of precipitation forecasts using deep generative models. In International conference on artificial intelligence and statistics (pp. 10555-10571). PMLR.

Price, I., Sanchez-Gonzalez, A., Alet, F., Ewalds, T., El-Kadi, A., Stott, J., ... & Willson, M. (2023). GenCast: Diffusion-based ensemble forecasting for medium-range weather. *arXiv preprint arXiv:2312.15796*.

Prinn, R. G. (2013). Development and application of earth system models. *Proceedings of the National Academy of Sciences*, *110*(supplement_1), 3673-3680.

Raissi, M., Perdikaris, P., & Karniadakis, G. E. (2019). Physics-informed neural networks: A deep learning framework for solving forward and inverse problems involving nonlinear partial differential equations. *Journal of Computational physics*, *378*, 686-707.

Rampal, N., Gibson, P. B., Sood, A., Stuart, S., Fauchereau, N. C., Brandolino, C., ... & Meyers, T. (2022). High-resolution downscaling with interpretable deep learning: Rainfall extremes over New Zealand. *Weather and Climate Extremes*, *38*, 100525.

Rasp, S., Pritchard, M. S., & Gentine, P. (2018). Deep learning to represent subgrid processes in climate models. *Proceedings of the National Academy of Sciences*, *115*(39), 9684-9689.

Rasp, S., Hoyer, S., Merose, A., Langmore, I., Battaglia, P., Russel, T., ... & Sha, F. (2023). WeatherBench 2: A benchmark for the next generation of data-driven global weather models. *arXiv preprint arXiv:2308.15560*.

Ravuri, S., Lenc, K., Willson, M., Kangin, D., Lam, R., Mirowski, P., ... & Mohamed, S. (2021). Skilful precipitation nowcasting using deep generative models of radar. *Nature*, *597*(7878), 672-677.

Reichstein, M., Bahn, M., Ciais, P., Frank, D., Mahecha, M. D., Seneviratne, S. I., ... & Wattenbach, M. (2013). Climate extremes and the carbon cycle. *Nature*, *500*(7462), 287-295.

Reichstein, M., Camps-Valls, G., Stevens, B., Jung, M., Denzler, J., Carvalhais, N., & Prabhat, F. (2019). Deep learning and process understanding for data-driven Earth system science. *Nature*, *566*(7743), 195-204.

Rezende, D., & Mohamed, S. (2015, June). Variational inference with normalizing flows. In International conference on machine learning (pp. 1530-1538). PMLR.

Richards, J., & Huser, R. (2022). Regression modelling of spatiotemporal extreme U.S. wildfires via partially-interpretable neural networks. *arXiv preprint arXiv:2208.07581*.

Rosenfeld, D., Sherwood, S., Wood, R., & Donner, L. (2014). Climate effects of aerosol-cloud interactions. *Science*, *343*(6169), 379-380.

Ross, A., Li, Z., Perezhogin, P., Fernandez-Granda, C., & Zanna, L. (2023). Benchmarking of machine learning ocean subgrid parameterizations in an idealized model. *Journal of Advances in Modeling Earth Systems*, *15*(1), e2022MS003258.

Sachindra, D. A., Ahmed, K., Rashid, M. M., Shahid, S., & Perera, B. J. C. (2018). Statistical downscaling of precipitation using machine learning techniques. Atmospheric research, 212, 240-258.

Sane, A., Reichl, B. G., Adcroft, A., & Zanna, L. (2023). Parameterizing Vertical Mixing Coefficients in the Ocean Surface Boundary Layer using Neural Networks. *Journal of Advances in Modeling Earth Systems*, 15, e2023MS003890.

Schär, C., Fuhrer, O., Arteaga, A., Ban, N., Charpilloz, C., Di Girolamo, S., ... & Wernli, H. (2020). Kilometer-scale climate models: Prospects and challenges. *Bulletin of the American Meteorological Society*, *101*(5), E567-E587.

Schneider, T., Lan, S., Stuart, A., & Teixeira, J. (2017). Earth system modeling 2.0: A blueprint for models that learn from observations and targeted high-resolution simulations. *Geophysical Research Letters*, *44*(24), 12-396.

Schulthess, T. C., Bauer, P., Wedi, N., Fuhrer, O., Hoefler, T., & Schär, C. (2018). Reflecting on the goal and baseline for exascale computing: a roadmap based on weather and climate simulations. *Computing in Science & Engineering*, *21*(1), 30-41.

Schultz, M. G., Betancourt, C., Gong, B., Kleinert, F., Langguth, M., Leufen, L. H., ... & Stadtler, S. (2021). Can deep learning beat numerical weather prediction?. *Philosophical Transactions of the Royal Society A*, *379*(2194), 20200097.

Serifi, A., Günther, T., & Ban, N. (2021). Spatio-temporal downscaling of climate data using convolutional and error-predicting neural networks. Frontiers in Climate, 3, 656479.

Sønderby, C. K., Espeholt, L., Heek, J., Dehghani, M., Oliver, A., Salimans, T., ... & Kalchbrenner, N. (2020). Metnet: A neural weather model for precipitation forecasting. *arXiv preprint arXiv:2003.12140*.

Stengel, K., Glaws, A., Hettinger, D., & King, R. N. (2020). Adversarial super-resolution of climatological wind and solar data. *Proceedings of the National Academy of Sciences*, *117*(29), 16805-16815.

Stevens, B., Acquistapace, C., Hansen, A., Heinze, R., Klinger, C., Klocke, D., ... & ZÄNGL, G. (2020). The added value of large-eddy and storm-resolving models for simulating clouds and precipitation. *Journal of the Meteorological Society of Japan. Ser. II*, *98*(2), 395-435.

Tipping, M. (1999). The relevance vector machine. *Advances in neural information processing systems*, *12*.

Vandal, T., Kodra, E., & Ganguly, A. R. (2019). Intercomparison of machine learning methods for statistical downscaling: the case of daily and extreme precipitation. *Theoretical and Applied Climatology*, *137*, 557-570.

Wang, Z., Chen, J., & Hoi, S. C. (2020). Deep learning for image super-resolution: A survey. *IEEE transactions on pattern analysis and machine intelligence*, *43*(10), 3365-3387.

Wang, R., Walters, R., & Yu, R. (2022, June). Approximately equivariant networks for imperfectly symmetric dynamics. In *International Conference on Machine Learning* (pp. 23078-23091). PMLR.

Watson, C. D., Wang, C., Lynar, T., & Weldemariam, K. (2020). Investigating two super-resolution methods for downscaling precipitation: ESRGAN and CAR. arXiv preprint arXiv:2012.01233.

Watson-Parris, D., Bellouin, N., Deaconu, L. T., Schutgens, N. A., Yoshioka, M., Regayre, L. A., ... & Stier, P. (2020). Constraining uncertainty in aerosol direct forcing. *Geophysical Research Letters*, *47*(9), e2020GL087141.

Willard, J., Jia, X., Xu, S., Steinbach, M., & Kumar, V. (2020). Integrating physics-based modeling with machine learning: A survey. *arXiv preprint arXiv:2003.04919*, *1*(1), 1-34.

Zanetta, F., Nerini, D., Beucler, T., & Liniger, M. A. (2023). Physics-constrained deep learning postprocessing of temperature and humidity. *Artificial Intelligence for the Earth Systems*.